\documentclass[a4paper,twocolumn]{esapub}
\usepackage{times}
\usepackage{natbib}
\usepackage{graphicx}

\title{Modelling flux tube dynamics}
\author{S{\o}ren Bertil Fabricius Dorch}
\affil{The Institute for Solar Physics of the Royal Swedish Academy of Sciences,
Stockholm Center for Physics, Astronomy and Biotechnology, 
SE-10691 Stockholm, Sweden}
\affil{Copenhagen University Observatory, 
the Niels Bohr Institute for Astronomy, Physics and Geophysics, 
Juliane Maries Vej 30, DK-2100 Copenhagen {\O}, Denmark}

\begin{document}


\vspace{-2.0cm}

\maketitle

\begin{abstract}
Over the last few years, numerical models of the behavior of solar magnetic
flux tubes have gone from using methods that were essentially one-dimensional
(i.e.\ the thin flux tube approximation), over more or less idealized 
two-dimensional simulations, to becoming ever more realistic 
three-dimensional case studies. Along the way a lot of new knowledge has been
picked up as to the e.g.\ the likely topology of the flux tubes, and the 
instabilities that they are subjected to etc. Within the context of what one
could call the ``flux tube solar dynamo paradigm,'' I will discuss recent
results of efforts to study buoyant magnetic flux tubes ascending from deep
below the photosphere, before they emerge in active regions and interact 
with the field in the overlying atmosphere (cf.\ the contributions by
Boris Gudiksen and {\AA}ke Nordlund): i.e.\ I am not addressing the flux tubes
associated with magnetic bright points, which possibly are generated by a
small-scale dynamo operating in the solar photosphere (cf.\ the
contribution by Bob Stein). The presented efforts are 
numerical MHD simulations of twisted flux ropes and loops,
interacting with rotation and convection. Ultimately the magnetic surface
signatures of these simulations, when compared to observations, constraints
the dynamo processes that are responsible for the generation
of the flux ropes in the first place. 
Along with these new results several questions pop up
(both old and new ones), regarding the nature of flux tubes and 
consequently of the solar dynamo.
\end{abstract}

\section{Introduction}

Buoyant magnetic flux tubes are an essential part 
of the framework of the current theories of dynamo action  
in both the Sun and solar-like stars: 
it is believed that when formed near the bottom of the 
convection zone (CZ), by a combination of rotation and 
turbulent convection, toroidal flux tubes buoyantly
ascend in the form of tubular $\Omega$-shaped loops. Rising under the
influence of rotational forces, they finally emerge
after a few months
as slightly asymmetric and tilted bipolar magnetic regions at the surface.
Many models of buoyant magnetic flux tubes are
based on the thin flux tube approximation \citep{Spruit1981}
that treats the tubes as strings, much thinner than e.g.\ the local pressure
scale height, moving subjected the Coriolis and drag forces. 
This essentially 1-d approximation is consistent with
the observations when used to study the latitudes of emergence, tilt angles, 
and the tilt-scatter  
of bipolar magnetic regions on the Sun, but only if the initial
field strength of the tubes are of the order of 10 times the convection
equipartition value near the bottom of the CZ
\citep{DSilva+Choudhuri93,Fan+ea94,Caligari+ea95}. A major problem in dynamo 
theory is to understand how
the field strength can become so high, corresponding to an energy density 
100 times larger than that of the available kinetic energy.


However, as buoyant flux tubes rise they expand and the assumption that they
are thin breaks down some 20 Mm below the solar surface. Traditionally,
the latter fact is taken as the main reason for ``going into higher 
dimensions'',
i.e.\ for submitting to 2-d (actually 2.5-d) and 3-d models.
So far, a lot of questions still remain unanswered, 
e.g.\ whether the quasi-steady state topology that the 
flux ropes reach in the later phase of their rise in 2-d simulations 
(e.g.\ Emonet \& Moreno-Insertis 1998 and Dorch et al.\ 1999) 
is stable towards
perturbations from the surroundings, and whether the results found
for 3-d flux ropes moving in a 1-d average static
stratification, at all are valid in the more realistic case. 
Hence when attempting to review the status of buoyant magnetic
flux tube models, there are several questions (Q's) that one may ask.
Below are a few examples:
~\\
{{\bf Q1:} In general the tubes may be twisted, but how much twist is 
 needed and warranted?}\\ 
{{\bf Q2:} How does the tube's twist evolve as they ascend: do they e.g.\ kink
 due to an increasing degree of twist?}\\
{{\bf Q3:} Are there other instabilities besides the magnetic 
 Rayleigh-Taylor (R-T) and kink instabilities?}\\
{{\bf Q4:} How do the tubes interact with the flows within the
 CZ and at the surface?}\\
{{\bf Q5:} What happens to the less buoyant magnetic subsurface structure as 
 the tubes rise (the wake)?}\\
{{\bf Q6:} What happens when the ropes become thick, i.e.\ large comparable to
 the local pressure scale height?}\\
{{\bf Q7:} What happens at emergence? Does the field topology change (how)?}\\ 
{{\bf Q8:} How does the twist arise? What is the appropriate initial condition?}

In the following I will try to answer some of
the above Q's by reviewing some of the 
main results of 2-d and 3-d models of buoyant magnetic flux tubes with 
emphasis on the most recent results from the turn of this millennium.

\section{Reference models}

For my discussion of the various aspects of flux tube models in the following,
I have made a number of 2-d and 3-d simulations for reference. These
are simulations for several values of 
the characteristic field line pitch angle $\psi_{\rm R}$. 

The 2-d reference simulations lack convection and the flux tubes move in 
a polytropic atmosphere (dynamic), while the 3-d
reference models retain full hydrodynamic convection with solar-like
super-granulation and down-flows (see the Appendix at the end of this paper).

To identify the models I use a terminology where the models are referred to
as e.g.\ 3D25 for a 3-d model where the flux tube initially is twisted
corresponding to a pitch angle of $\Psi_{\rm R} = 25^{\rm o}$. A 2-d 
model with the same pitch angle is referred to as 2D25.
All the reference models have 
plasma-$\beta$'s at the onset of the model equal to  $\beta_0=100$.
The set-up of the reference models are discussed in more detail in the
Appendix, but the 
initial twist of the flux tubes is given by
\begin{equation}
 {\rm B}_z = {\rm B}_{\rm 0} e^{-(r/{\rm R})^2} {\rm ~and~}
 {\rm B}_{\phi} = \alpha r / {\rm R}~ {\rm B}_z, \label{initial.eq}
\end{equation}
where ${\rm B}_z$ is the parallel
and ${\rm B}_{\phi}$ the transversal component 
of the magnetic field with respect to the tube's main axis. 
${\rm B}_{\rm 0}$ is the amplitude of the field, R the radius,
and $\alpha$ is the field line pitch parameter. 
The tubes are twisted, with a maximum pitch angle
of $\psi_{\rm R}= \arctan (\alpha)$, hence ``rope'' is a better word to
describe the state of the field lines than ``tube''.
In the following I set the initial 
radius to ${\rm R}_0 = 0.177~ {\rm H}_{\rm P0}$, i.e.\ the tubes are
non-thin. 

The wavelength $\lambda$ of the flux rope is equal to the horizontal size of the
domain $\lambda = 3.2~ {\rm H}_{\rm P0}$ 
at the initial position of the rope where the pressure scale height is
${\rm H}_{\rm P0} = 78.3$ Mm.
Then the flux rope is not undular Parker-unstable even though the 
stratification admits this instability for ropes longer 
than a critical wavelength of $\sim 12~ {\rm H}_{\rm P0}$ 
\citep{Spruit+Ballegooijen82}.

Primarily the behavior of the flux tubes during 
the initial and rise phases are discussed. When the ropes approach the upper
boundary of the CZ,
they enter an ``emergence phase'' slightly beyond the scope of this review 
(but see e.g.\ the contributions by Gudiksen and Nordlund, ibid).

\begin{figure}[!htb]
\centering
\vbox{\includegraphics*[width=8cm, height=7cm] {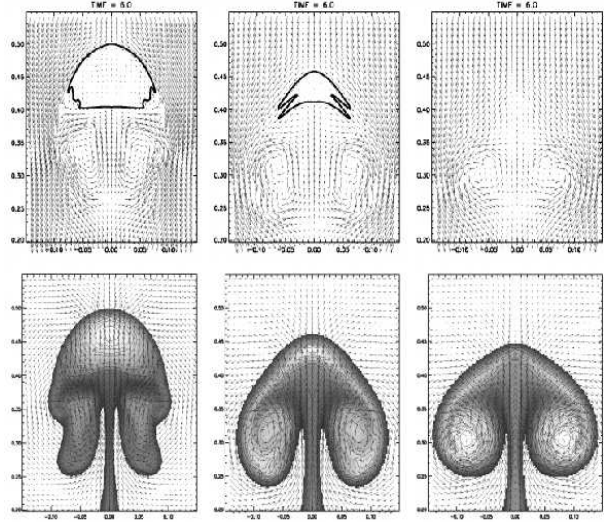} }
\caption{Flux tube disrupted by the magnetic Rayleigh-Taylor 
instability; from a 2-d simulation by \cite{Emonet+Moreno98}:
The typical pitch angle was set to $7^{\rm o}$ (left), $2^{\rm o}.5$ 
(middle) and $0^{\rm o}$ (right). Shown are equipartition curves (top row),
field strength (bottom row), and flow pattern (tiny arrows).} 
\label{fig1a}
\end{figure}

\section{Twisted tubes}

2-d simulations of flux tube cross-sections, not 
relying on the assumption that the
flux tubes are thin, have shown that purely cylindrical tubes are quickly
disrupted by the magnetic R-T instability, 
rendering them unlikely to reach the surface
because they loose their buoyancy and cohesion
\citep{Schussler1979,Tsinganos1980,Emonet+Moreno98,Dorch+Nordlund98,Krall+ea98,
Wissink+ea00}.
\begin{figure*}[!htb]
\centering
\vbox{\includegraphics*[width=4.5cm, height=4.5cm] {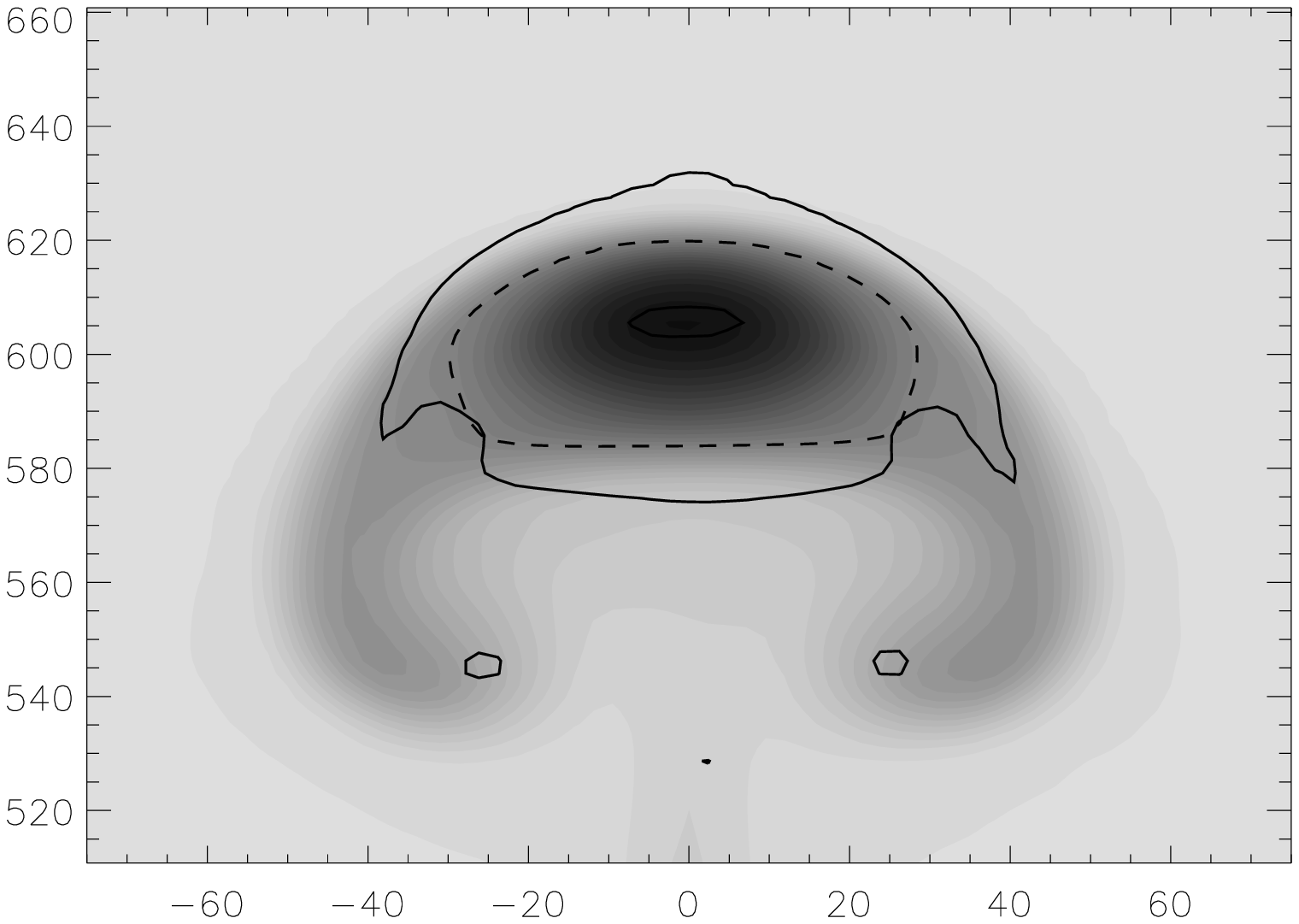} 
      \hspace{-0.5cm}
      \includegraphics*[width=4.5cm, height=4.5cm] {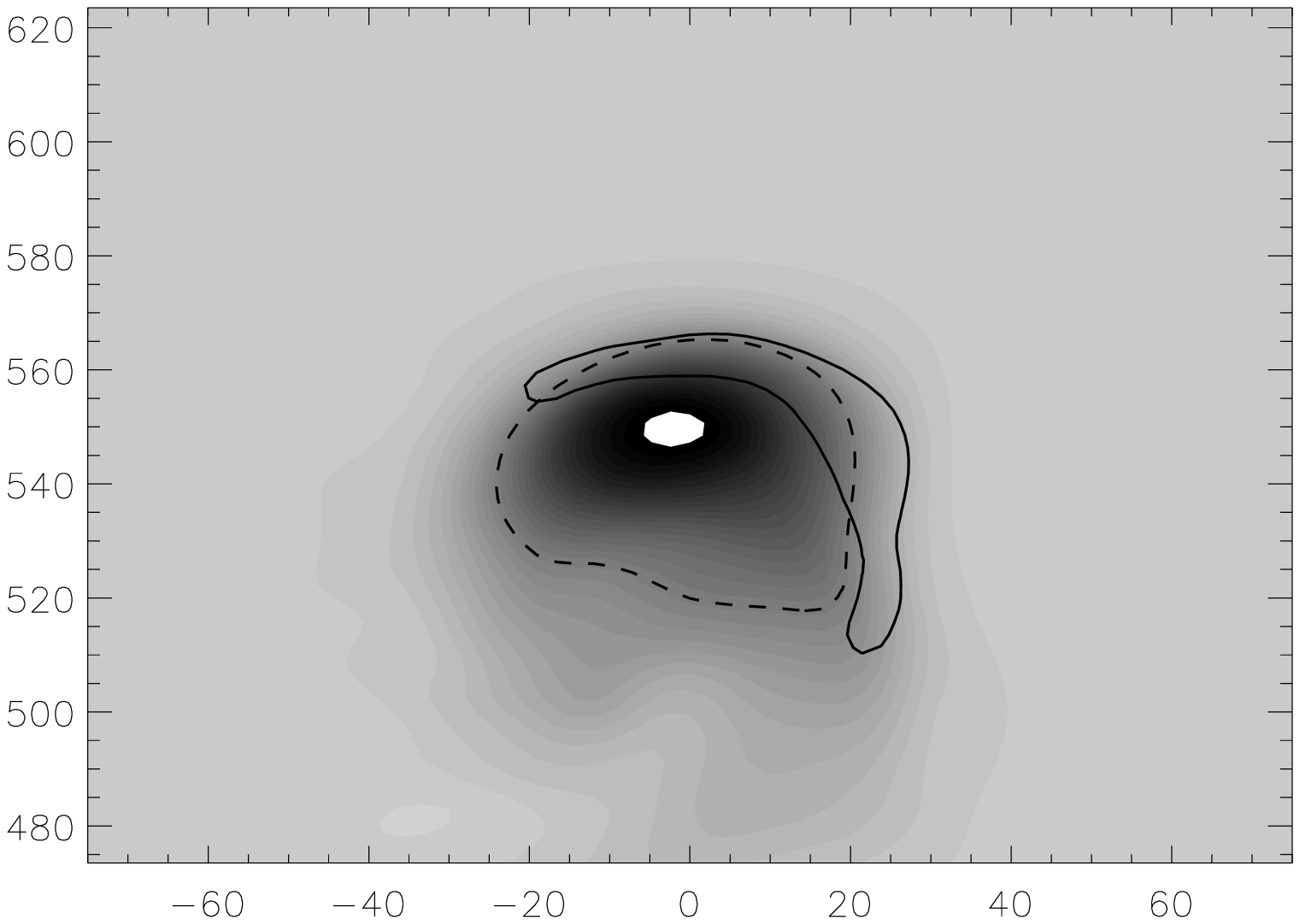} 
      \hspace{-0.5cm}
      \includegraphics*[width=4.5cm, height=4.5cm] {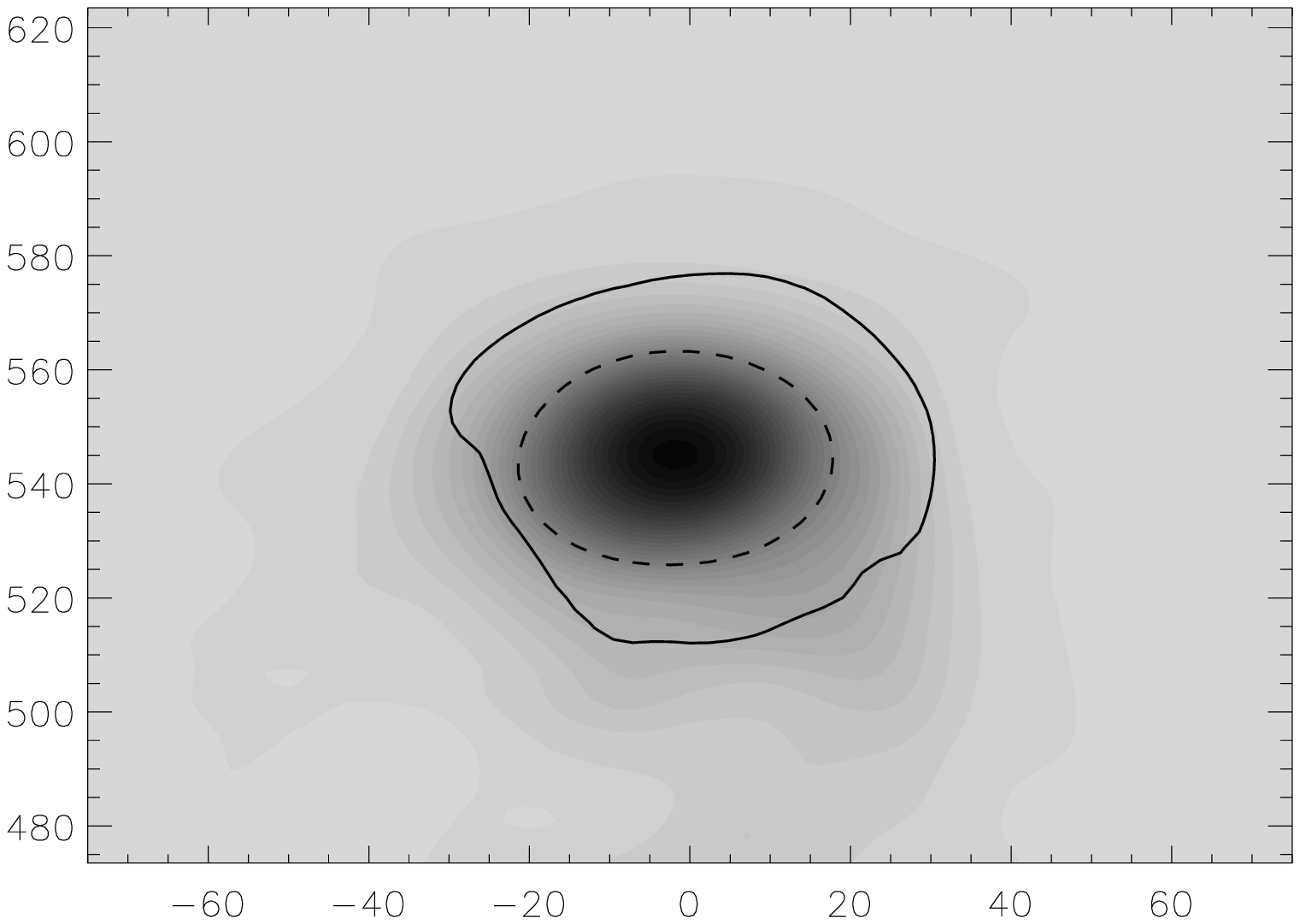} }
\caption{Images showing
${\rm B}_z$ at time $67.7~ \tau_{\rm A}$ for2D30 (left), and at time  
$35.6~ \tau_{\rm A}$ for 3D25 and 3D45
(center and right). Shown by contours 
are the equipartition curves (full curves). 
Also shown are HWHM-contours of ${\rm B}_z$ (dashed).
Only a subsection of the full computational box is shown.} 
\label{fig3}
\end{figure*}
This apparent disruption of the tubes is a result of the too simple topology
assumed for their magnetic field: 
rather, the magnetic field line tension present in a twisted flux 
{\em rope}
suppresses the R-T instability and hence prevents the 
flux structures from disintegrating as it has been demonstrated  
by e.g.\ Emonet \& Moreno-Insertis (1998):
an expression similar to Eq.\ (1) has been applied in both
2-d and 3-d simulations and it has been shown that for these cases the 
R-T instability is inhibited
if the degree of twist is sufficiently high as shown by
\cite{Emonet+Moreno98}: the corresponding critical value of the 
field line pitch angle $\Psi_c$ is approximately
determined by equating the energy
density of the transversal (or azimuthal) field component equal to
the ram pressure of the flow relative to the flux rope
(incidentally, several approaches yields the same expression for $\Psi_c$).  
This kind of twist-topology
is simple but similar to the relaxed state of a flux rope with 
a more complicated topology used in \cite{Dorch+Nordlund98}.
\cite{Emonet+Moreno98} finds a critical pitch angle on the order of
$arcsin (\sqrt{{\rm R}/{\rm H}_{\rm P0}})$, which typically is
$10^{\rm o}$ for thin flux ropes. For thicker 
flux ropes, like the ones I discuss here, the value of $\Psi_c$ increases to 
approximately $25^{\rm o}$. 

The boundary of a flux rope can be defined by the equipartition curve
(the locations where
the transversal magnetic field is in equipartition with the kinetic energy
relative to the rise of the rope).
Fig \ref{fig1a}.\ shows equipartition curves for 2-d models by
\cite{Emonet+Moreno98}, corresponding to  three values of the pitch angle:
only the most twisted ropes withstand the exterior
pressure fluctuations, and
the transversal field is strong enough to protect the buoyant core of the
flux rope.  

The results of my 2-d reference simulations agree perfectly with the  
2-d studies mentioned above: the ropes are disrupted by the R-T instability
unless their pitch angle exceeds a critical value. As the ropes
rise and expand they enter a ``terminal rise phase'' where they ascent with a 
constant speed---while oscillating due to their differential buoyancy---before 
reaching the upper computational boundary (which is open). 
In this rise-phase the speed is the so-called terminal velocity $v_t$ 
determined by the balance between buoyancy and drag. 
In reference simulation 2D30---see Fig.\ \ref{fig3} (left)---I get
$v_t \approx 0.1~ v_{\rm A0}$ (with $v_{\rm A0}$ being the
Alfv\'{e}n speed $v_{\rm A}$ at the initial position).
The largest Mach number reached during the entire rise 
is $8.5\times 10^{-2}$, so that the motions are completely sub-sonic.
The characteristic time scale of the ascent then becomes $\tau_{\rm rise}
= {\rm H}_{\rm P0}/v_t \approx 50~ \tau_{\rm A}$ with $\tau_{\rm A} = 
{\rm R}_0/v_{\rm A0}$.
From the 2-d reference
simulations I find that a minimum pitch angle in the range $25^{\rm o}$ -- 
$30^{\rm o}$ is required for the ropes to be unaffected by the 
R-T instability, consistent with the result of \cite{Emonet+Moreno98}. 

Consider a 3-d reference simulation with $\Psi_{\rm R} = 25^{\rm o}$
(3D25): in that case the pitch is far to low to excite
the kink instability (see next Section), but we expect to 
re-find the R-T instability in addition to the action of the 
convective flows.
Fig \ref{fig3}.\ (center and right) 
shows cross-sectional averages of the magnetic field in
3D25 and 3D45 which may be compared to the corresponding image from 2D30
(note the difference in vertical offset). In the 2-d case, the 
horizontal and vertical sizes of the ropes do not increase equally fast.
This effect is primarily because of the differential
buoyancy that squeezes the ropes in the vertical direction, and by flux
conservation, expands them horizontally. The effect is less apparent in
3-d, where the ropes are more circular (see Fig \ref{fig3})
when defined by the HWHM-boundary (i.e.\ the half-width at half-maximum
of the parallel field component). 
The equipartition curve, however, loses its 2-d mirror symmetry.
For a less twisted rope (e.g.\ 3D25 in Fig.\ \ref{fig3}),  
as the rope rises and the R-T
instability sets in, the flux rope disrupts and eventually fills the bulk 
of the CZ. At late stages the magnetic field forms network
patches at the surface cell boundaries, while the pumping effect
\citep{Dorch+Nordlund01} transports
the weakest flux down towards the bottom of the CZ.

\subsection{Rotation and $\Omega$-loops}

It has been suggested that the value of the critical 
degree of twist needed to prevent the R-T instability may be unrealistically
high in the 2-d case, and a smaller twist may be sufficient in the case of 
sinusoidal 3-d magnetic flux loops 
as in the simulations: \cite{Abbett+ea00} used their anelastic MHD code to
study the buoyant rise of twisted flux loops and found that the greater
the curvature of the loop at its apex, the smaller the critical pitch angle.
Moreover, \cite{Abbett+ea01} found that including solar-like
rotation and a Coriolis
force when solving the MHD equations lead to less fragmentation, even
of initially un-twisted buoyant flux tubes. However, the limit on $\Psi_c$
is not known. 

\subsection{Sigmoids}

\begin{figure*}[!htb]
\centering
\hbox{ 
 \includegraphics*[width=8cm, height=3.75cm] {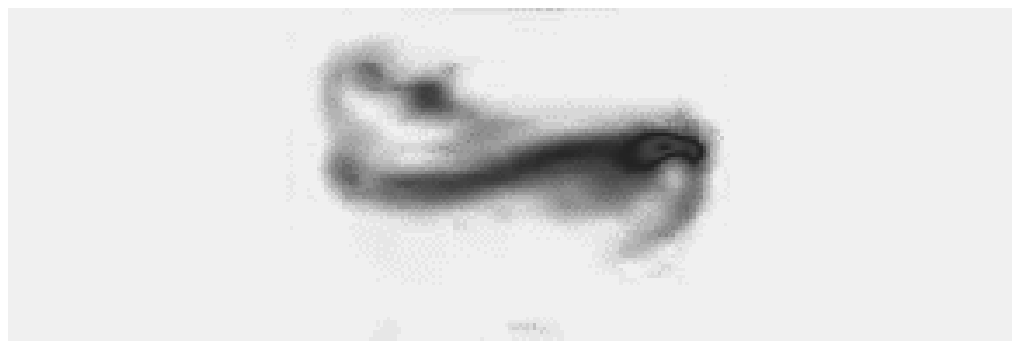} 
 \includegraphics*[width=8cm, height=3.75cm] {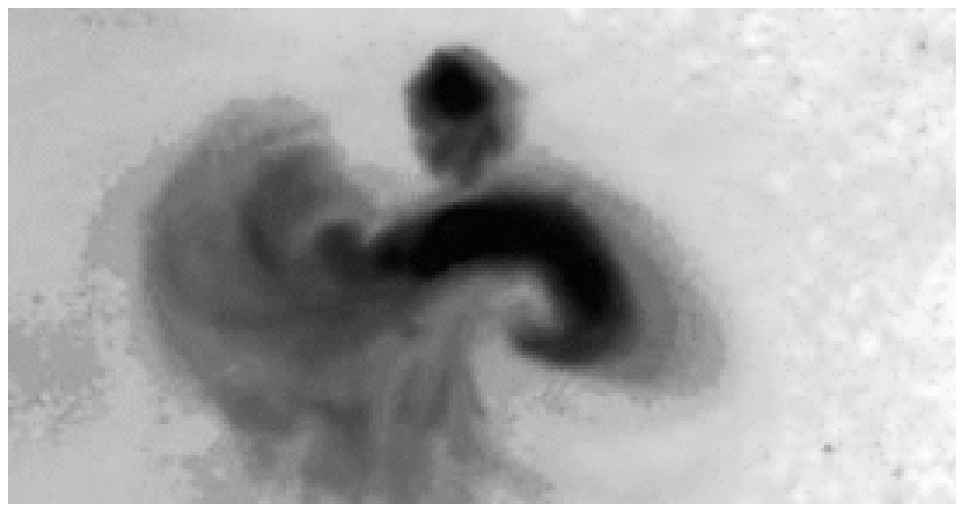} }
\caption{S-shaped structure of emerging magnetic flux rope from a
simulation by \cite{Dorch+ea99} (left), and a sigmoid in X-ray from 
the Yohkoh satellite (right).}
\label{sigmoid.fig}
\end{figure*}

Recently, 3-d simulations of buoyant twisted flux loops have confirmed that
several of the results found in the 2-d simulations carry over to the more
realistic 3-d scenario 
\citep{Matsumoto+ea98,Dorch+ea99,Abbett+ea00,Magara+Longcope01}.
From the 3-d results it is apparent that the apex 
cross-sections of twisted flux loops are well described by the 2-d studies.
Furthermore,
the S-shaped structure of a twisted flux loop as it emerges through the upper 
computational boundary is qualitative similar to the sigmoidal structures 
observed in EUV and soft X-ray by the Yohkoh and SoHO satellites, see
Fig.\ \ref{sigmoid.fig} 
\citep{Canfield+ea99,Sterling+ea00}.

\begin{figure}[!htb]
\centering
\vbox{ \includegraphics*[width=8cm, height=18cm] {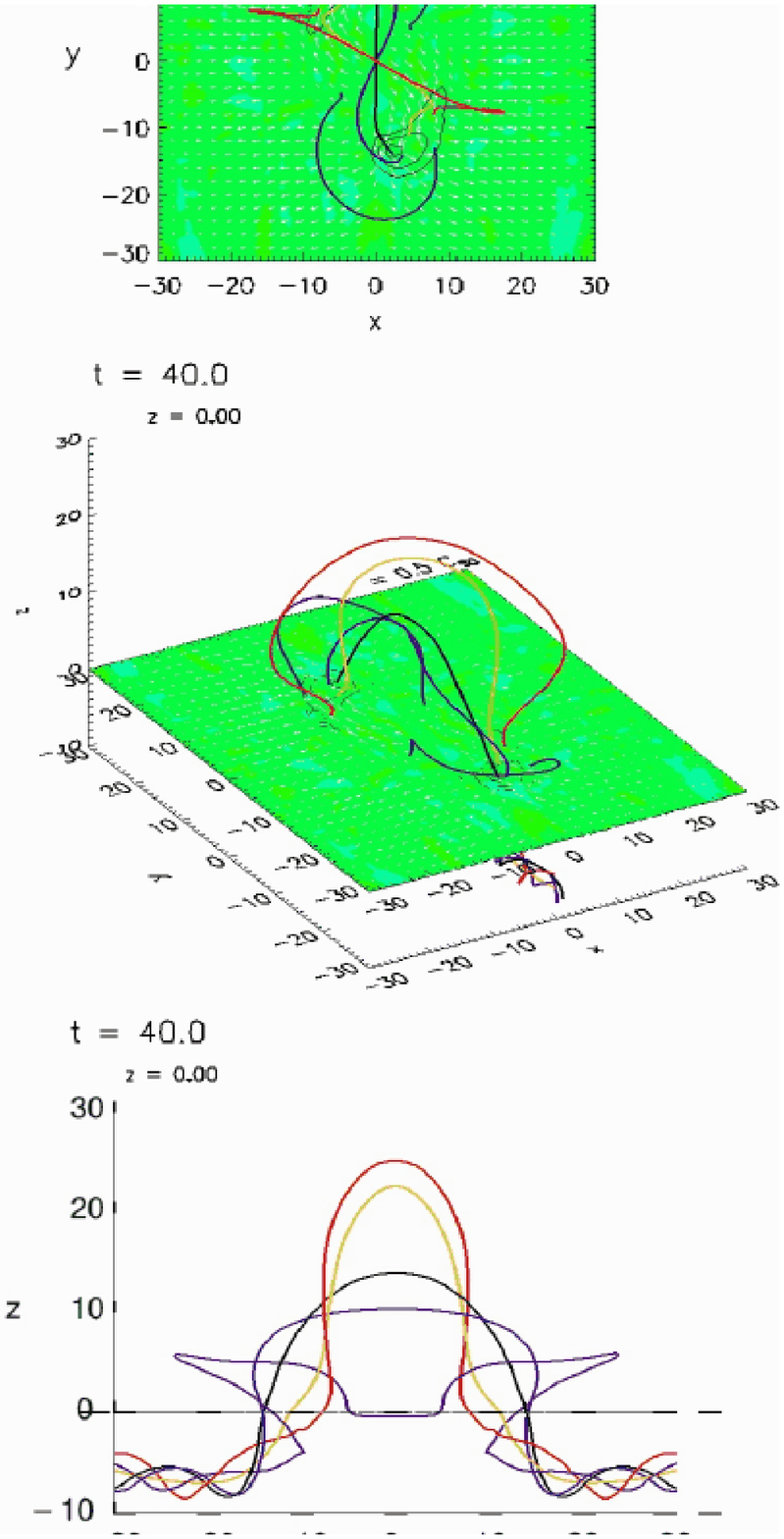} }
\caption{Snapshot of emerging twisted flux rope (dark field lines,
 horizontal velocity field as white arrows, vertical velocity in gray 
 scale, and vertical magnetic field component as contours): The darkest
 magnetic field line is the central axis of the flux rope.
 Figure provided by T.\ Magara, e.g.\ \cite{Magara+Longcope01}.}
\label{magara.fig}
\end{figure}

An example of such 3-d simulations are provided by \cite{Magara+Longcope01}
who performed ideal MHD simulations of a flux rope emerging from a convection 
zone through a photosphere into an overlying corona:
initially the flux rope 
is horizontal and in mechanical equilibrium, but when subjected to the undular
Parker instability it is forced to rise and break through the surface,
because of the buoyancy enhancing downflows along the its field lines.
A similar study was performed by \cite{Fan2001}.
The result is that first a magnetic bipole is formed at the surface by 
the oppositely directed vertical field lines of the looping flux rope.
Secondly, upon emergence, a neutral line and an S-shaped (sigmoidal) 
structure is formed by the flux rope's central (inner) field lines
see Fig.\ \ref{magara.fig}. The outer
field lines expand and form an S with the opposite sign of the inner one,
as well as an arcadial structure with an almost potential geometry.

\subsection{The flux rope's path}

\begin{figure*}[!htb]
\centering
\vbox{ \includegraphics*[width=14cm, height=7cm] {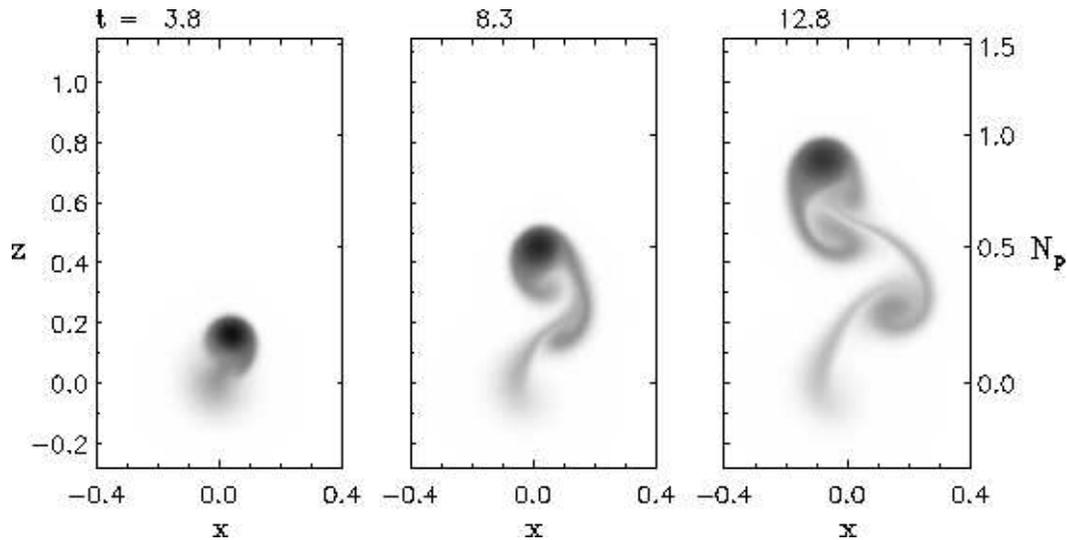} }
\caption{2-d simulation of a buoyant flux rope rising through a stratified
 atmosphere; the rope follows a zigzag trajectory. Figure provided by
 T.\ Emonet, e.g.\ \cite{Emonet+ea01}.}
\label{emonet.fig}
\end{figure*}

The naivistic view of a buoyant flux rope ascending 
nicely on a straight vertical (i.e.\ strictly speaking radial) path towards 
the surface can be
dismissed qualitatively; in the 3-d case with convection,
it is easy to imaging that the rope will be influenced by convective flows
and be pushed around, as we shall see below. However, even in 2-d, 
neglecting flows that is not connected with the rope's buoyant rise, a more 
subtle effect courses the ropes to have very complex trajectories
when the Reynolds number becomes sufficiently large: 
\cite{Emonet+ea01} found a zig-zag motion of the magnetic flux rope 
caused by the imbalance of vorticity in the flux tubes,
see Fig.\ \ref{emonet.fig}. 
Vorticity generated when flux ascend supply a lift force; it is by
this mechanism that weakly twisted flux ropes loose their updrift (negative
lift force from two vortex rolls). At high viscous Reynolds number the 
shredding of vortex rolls in the trailing Karman-like wave of the flux
introduces a vorticity-imbalance, resulting in an alternating lift force
shuffling the rising rope. Perhaps surprisingly, 
it turns out that the complicated path of the flux rope
can be described by a single non-dimensional number.  

\begin{figure*}[!htb]
\centering
\vbox{\includegraphics*[width=4.5cm, height=4.5cm] {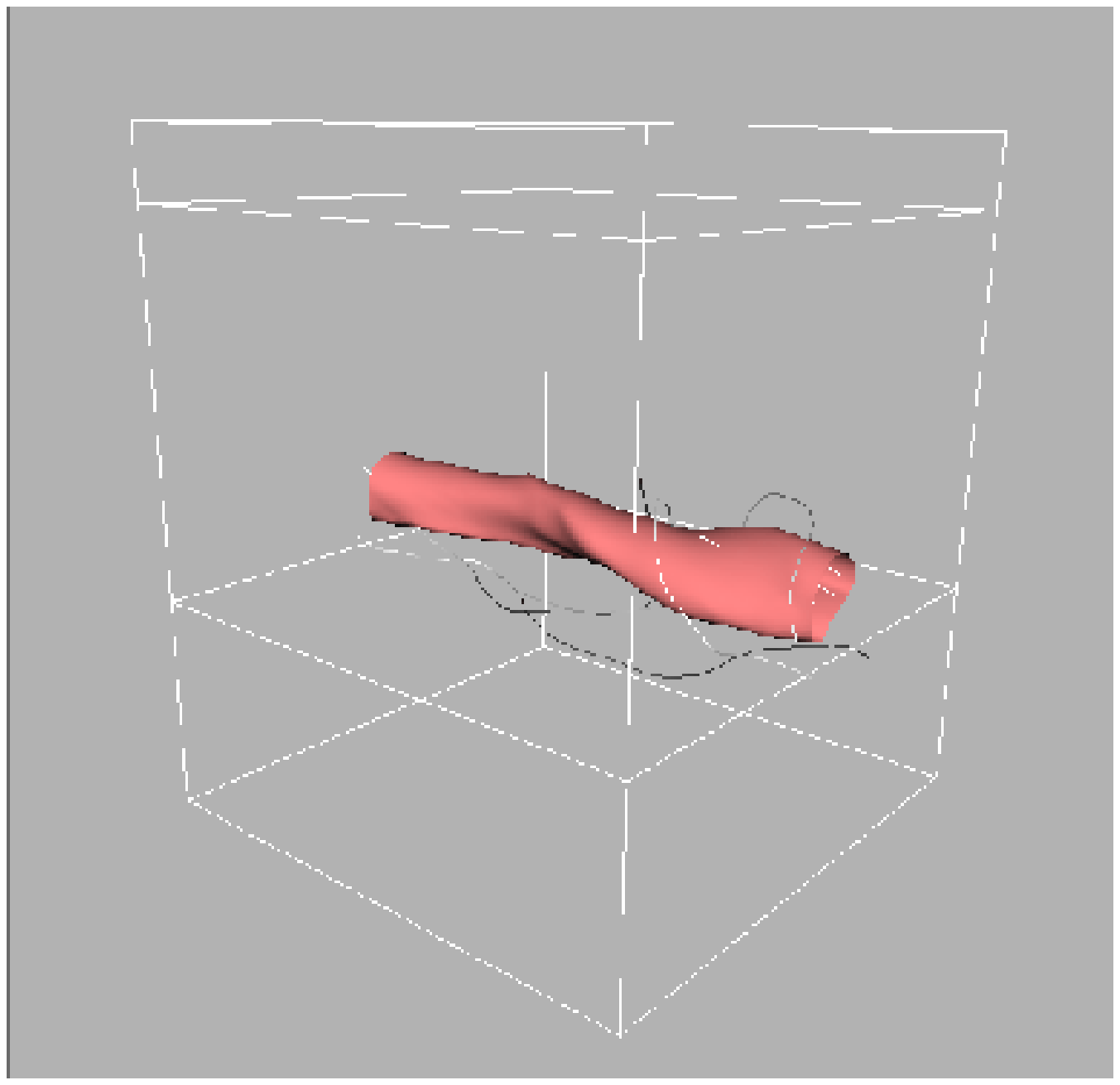} 
      \hspace{-0.5cm}
      \includegraphics*[width=4.5cm, height=4.5cm] {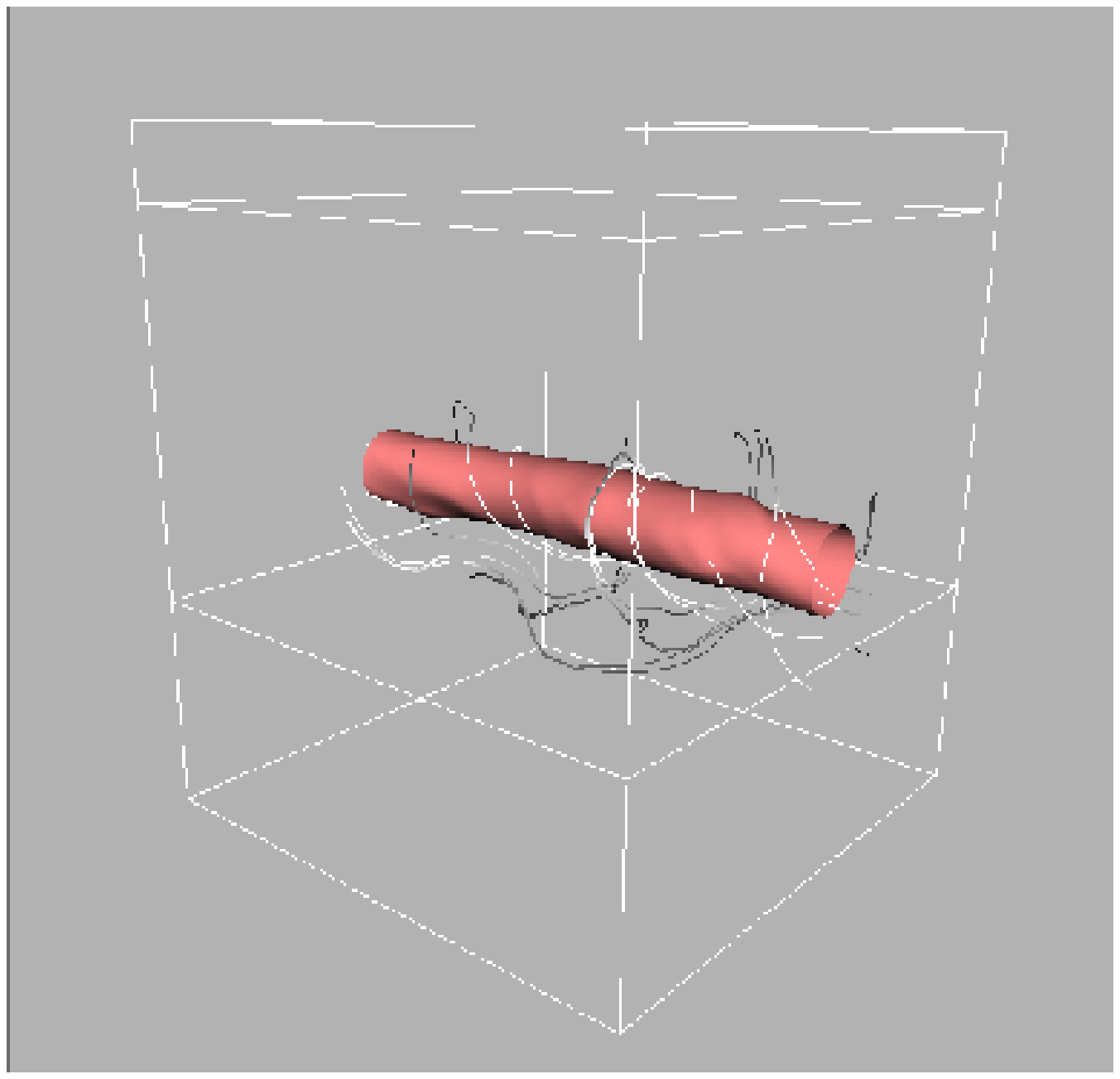} 
      \hspace{-0.5cm}
      \includegraphics*[width=4.5cm, height=4.5cm] {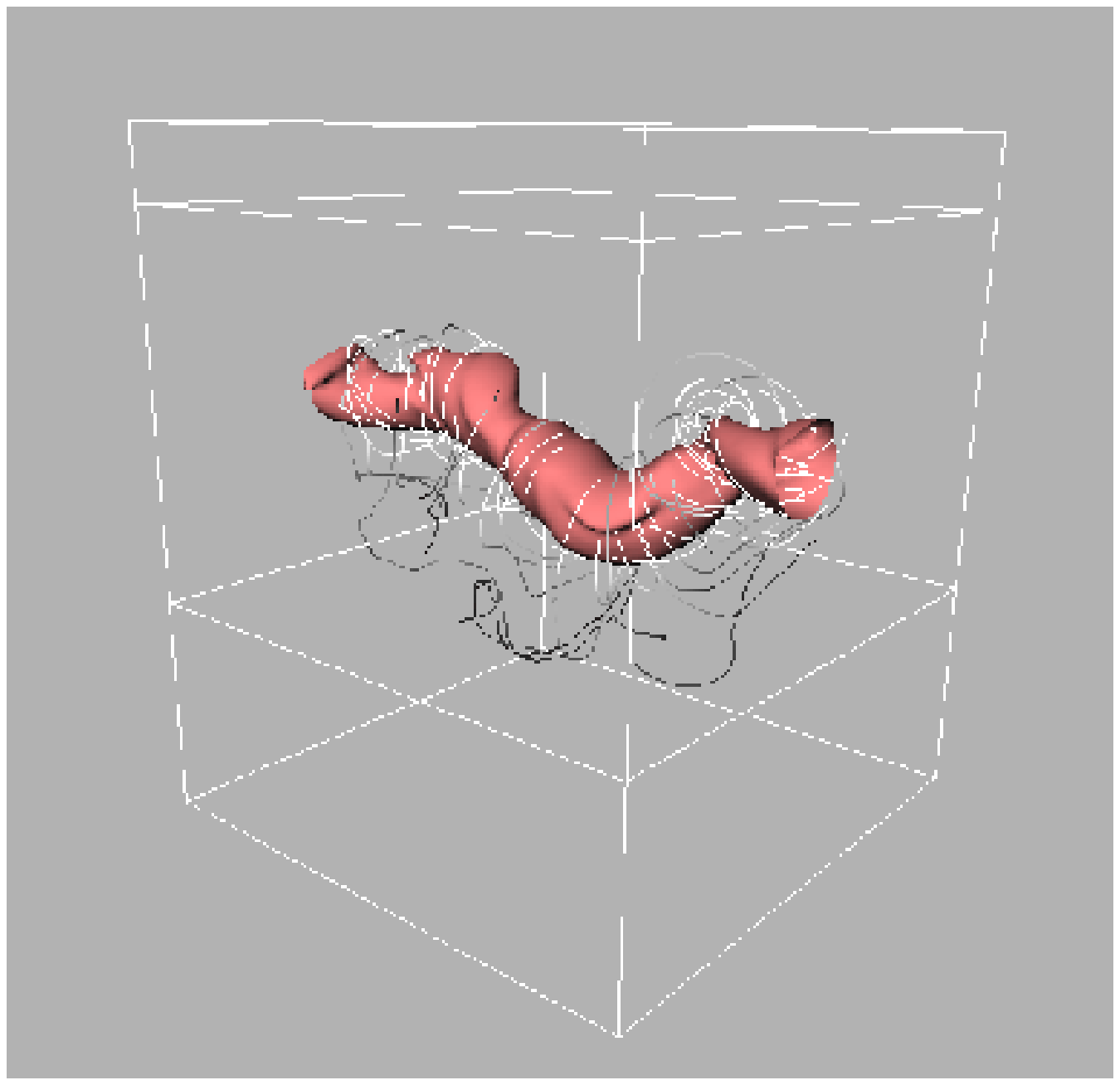} }
\caption{
Snapshots of magnetic field isosurfaces at time $t= 28~ \tau_{\rm A}$
for the simulations (left to right) 3D25 (weak twist, R-T unstable),
3D45 (stable rope) and 3D75 (strong twist, kink unstable rope):
the magnetic field isosurfaces outline the HWHM surfaces.} 
\label{fig4}
\end{figure*}

In the case 3D45, the degree of twist is small enough to 
prevent the onset of the kink instability
(the linear ideal growth rate vanishes for $\alpha = 1$, e.g.\ Fan et al.\ 
1999), yet it is large enough to prevent also the onset of the R-T instability:
Fig.\ \ref{fig4} illustrates the effect of these instabilities on the shape 
of the flux ropes.
Thus, in 3D45 the rope retains its cohesion without distorting its shape
by any of these two instabilities, 
and we may focus our attention on the effects of the convective flows on 
the rope.  

\begin{figure}[!htb]
\centering
\vbox{\includegraphics*[width=5.6cm,height=5cm]{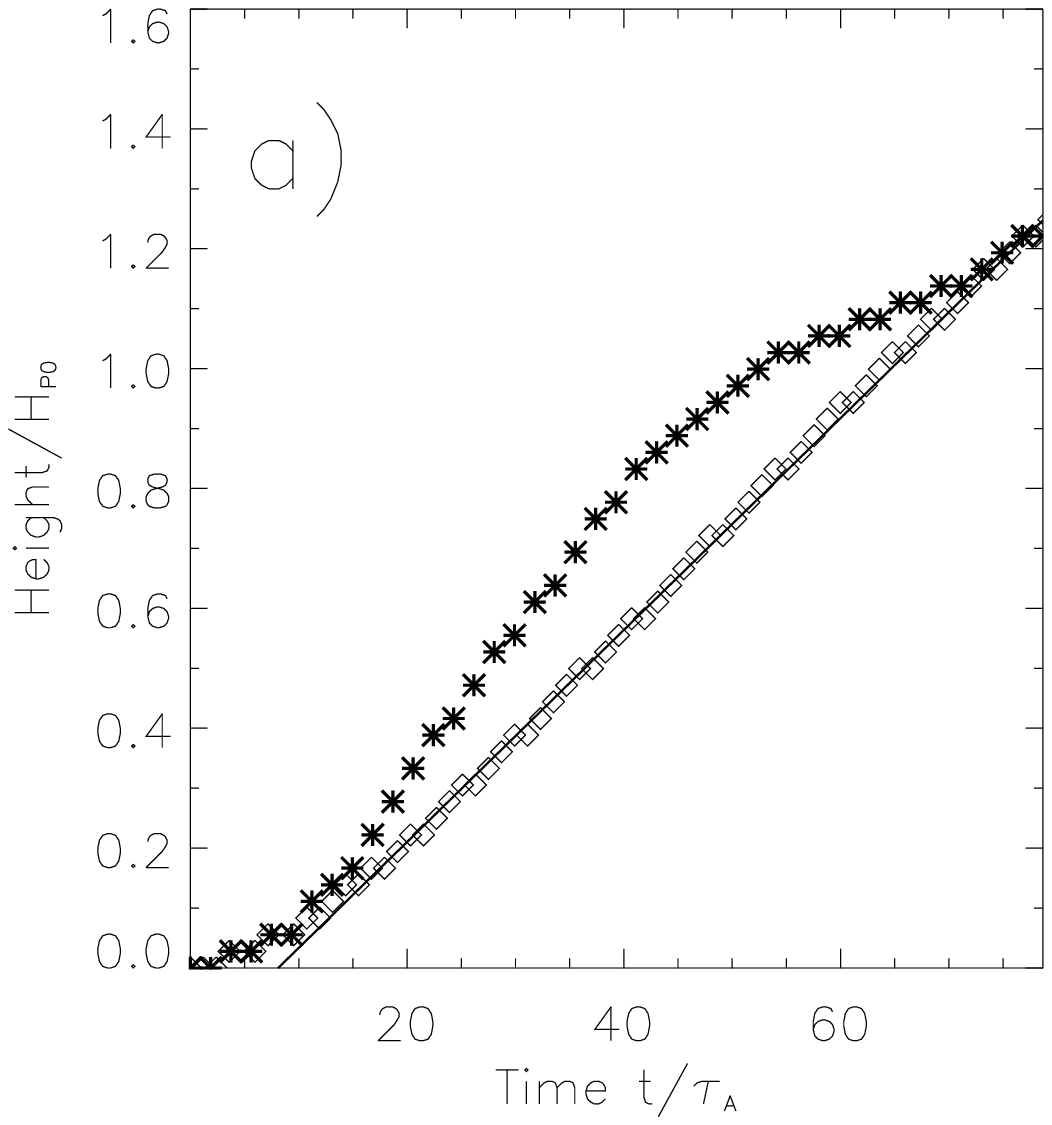}
      \hspace{-1.7cm}
      \includegraphics*[width=2.8cm,height=5cm]{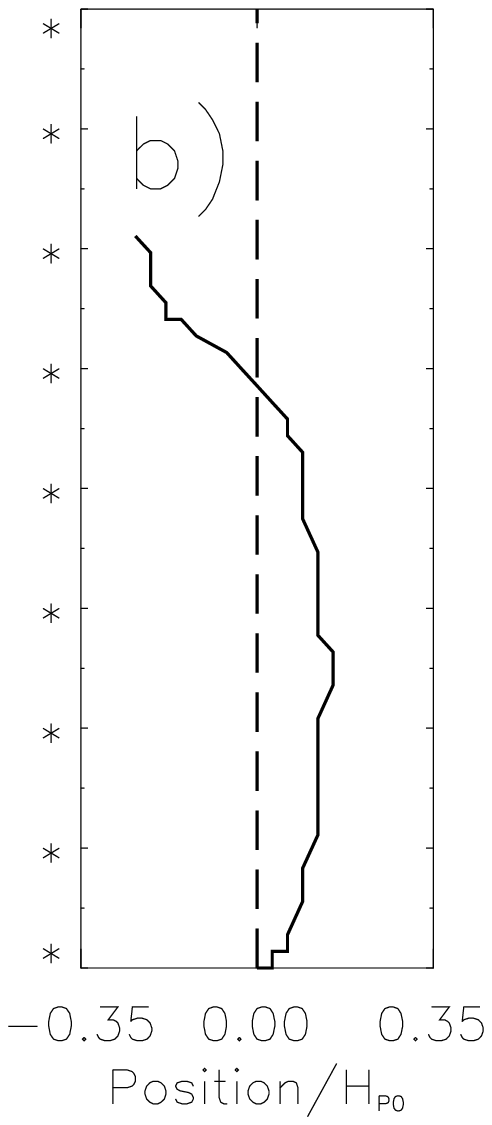}
}
\vspace{0.cm}
\caption{
 (a) Height of the flux rope as a function of time (3D45, stars and
  2D45, diamonds).
 The straight line corresponds to the average speed of $0.1~ v_{\rm A0}$ in 
 the rise phase.
 (b) Average drift of the 3D45 flux rope in the meridional plane.
} 
 \label{fig6}
\end{figure}

The initial position of the ropes in the CZ is significant for the 
subsequent detailed history of their rise: with the present
flows and location of the flux ropes, 
most parts of the ropes are incidentally
located inside or close to a convective updraft. 
Thus, the ascent of the ropes are influenced by this:
Fig \ref{fig6}.\ compares 3D45 to 2D45 
(i.e.\ to a convection-less reference simulation), 
and a simple analytic flux tube.
As the 3-d rope rises, convective flows perturb its motion, preventing it 
from entering a rise phase with a constant rise speed, as it indeed does
in 2D45 (see Fig \ref{fig6}a.) 
The rope remains straight and the maximum excursion of its axis, at the end of
the simulation, is $\sim 0.04~\lambda$.
With the chosen super-equipartition axial field strength, the main action of 
the large-scale convective flows is to push the rope both left and right of 
the central plane (Fig \ref{fig6}b.\ --- see also the movie on the 
CD-rom),
while the effect of the small-scale downdrafts 
is to locally deform its equipartition boundary.

The initial location within a general updraft region explains why 
the rise
speed of the rope is slightly greater than that of the reference simulation,
which reaches a terminal speed of $\sim 0.1~ v_{\rm A0}$.
The 3-d rope also expands more quickly than the rope in the 2-d simulation
(see Fig.\ \ref{rise.fig}), 
but its rate of expansion is closer to what is expected 
from an adiabatically expanding, non-stretching tube with constant flux.

\section{Kinking ropes}

Fan et al.\ (1999) report on 3-d non-convective simulations of 
buoyant kink-unstable flux ropes in a stratified atmosphere, and found that
a twisted flux rope becomes significantly unstable only when the 
pitch parameter
$\alpha$ is well above a critical ideal-MHD value $\alpha_c = 1$,
i.e.\ if $\psi_{\rm R} > 45^{\rm o}$: in the resistive-MHD 
case $\alpha_c$ increases slightly.
In reference simulation 3D75 a kink therefore develops, with 
a maximum ideal growth rate of $\Gamma_{\rm kink} \propto
\tau_{\rm A}^{-1}(\alpha^2 -1)$ (from Linton et al.\ 1998),
yielding a characteristic time scale of
$\Gamma_{\rm kink}^{-1} \approx 0.29~ \tau_{\rm A}$
significantly shorter than the rise time derived from the 
2-d reference simulation 2D75, see Fig.\ \ref{fig4}.
Diffusion decreases the growth rate further,
but the instability still sets in rather quickly, well before
the rope reaches the surface: the rope has plenty of time to kink
while rising.

I do not invoke a perturbation of the rope's mass or entropy to onset the 
kink,
as do e.g.\ Abbett et al.\ (2000) and others. Rather I let the rope 
be perturbed by the convective flows. The power of the convection
at the initial position of the rope is predominantly at long wavelengths
(100 Mm), and these wavelengths of the unstable
kink modes dominate the rope's evolution.

\begin{figure}[!htb]
\centering
\vbox{\includegraphics*[width=5cm,height=5cm]{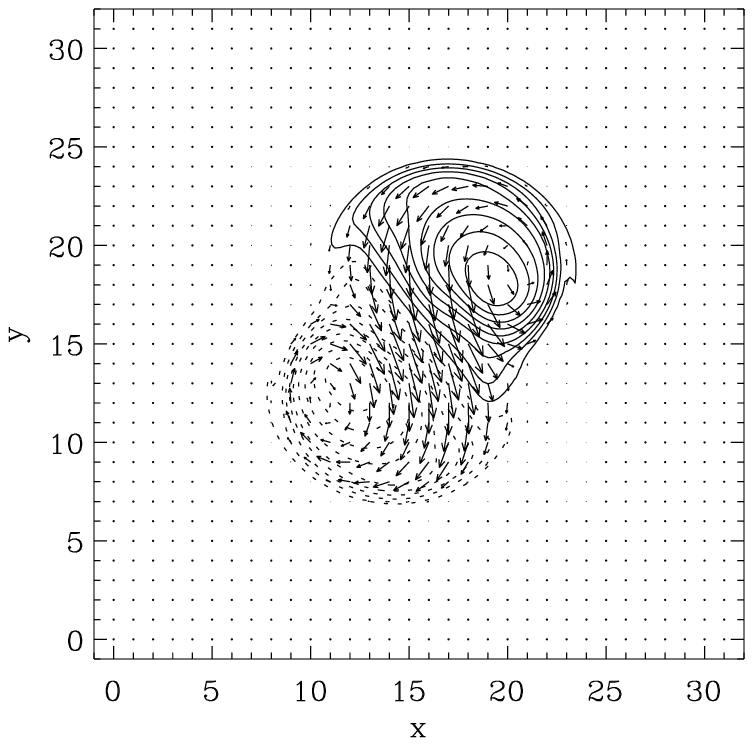}
}
\vspace{0.cm}
\caption{Simulation vector magnetogram of an emerging kinked rope
 showing the horizontal (arrows) and vertical magnetic field (contours), 
 adopted from \cite{Fan+ea99}. } 
 \label{fan.fig}
\end{figure}

As the rope kinks, it develops a significant curvature and a knotty shape
with portions dominated by vertical field components, see Fig.\ \ref{fig4}
(right). Upon emergence the bipole rotates and the neutral line between
the two main polarities are strongly sheared, see Fig.\ \ref{fan.fig}.
Hence, when the rope approaches the surface, it contains adjacent mixed
polarities in horizontal cross-sections (see the CD-rom movies); this is
reminiscent of the appearance of $\delta$-type sunspots with different 
polarities within a common umbra, and kinked flux
tubes are often quoted as possibly being responsible for this phenomenon.

\section{Loosing flux}

\cite{Petrovay+Moreno97} suggested that 
turbulent erosion of magnetic flux tubes may take place
due to the ``gnawing'' on a flux tube's field lines
by turbulent convection:
the flux tube is eroded by a thin current
sheet forming spontaneously within a diffusion time.

\begin{figure}[!htb]
\centering
\vbox{\includegraphics*[width=5.6cm,height=5cm]{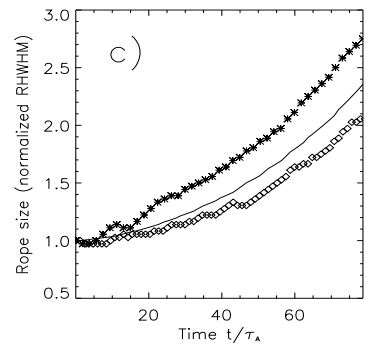}
}
\vspace{0.cm}
\caption{
  Expansion of the flux rope (3D45, stars and 2D45, diamonds) in
  units of its initial radius. 
  Also plotted is an analytical expression (solid line, see text).
} 
 \label{rise.fig}
\end{figure}

Fig \ref{rise.fig} shows the 3D45-rope's 
characteristic radius ${\rm R}_{\rm hwhm}$
along its axis (with the radius defined as the average HWHM). 
As the rope rises and expands, its magnetic field strength 
also decreases to approximately conserve the flux (the decrease found in 
3D45 is at a rate close but not identical to that expected for 
a flux-conserving tube).
The deviation can be attributed to the fact that, during its ascent, 
magnetic flux within the 3-d rope is lost to its surroundings.
\begin{figure}[!htb]
\centering
\vbox{\includegraphics*[width=8cm,height=6cm]{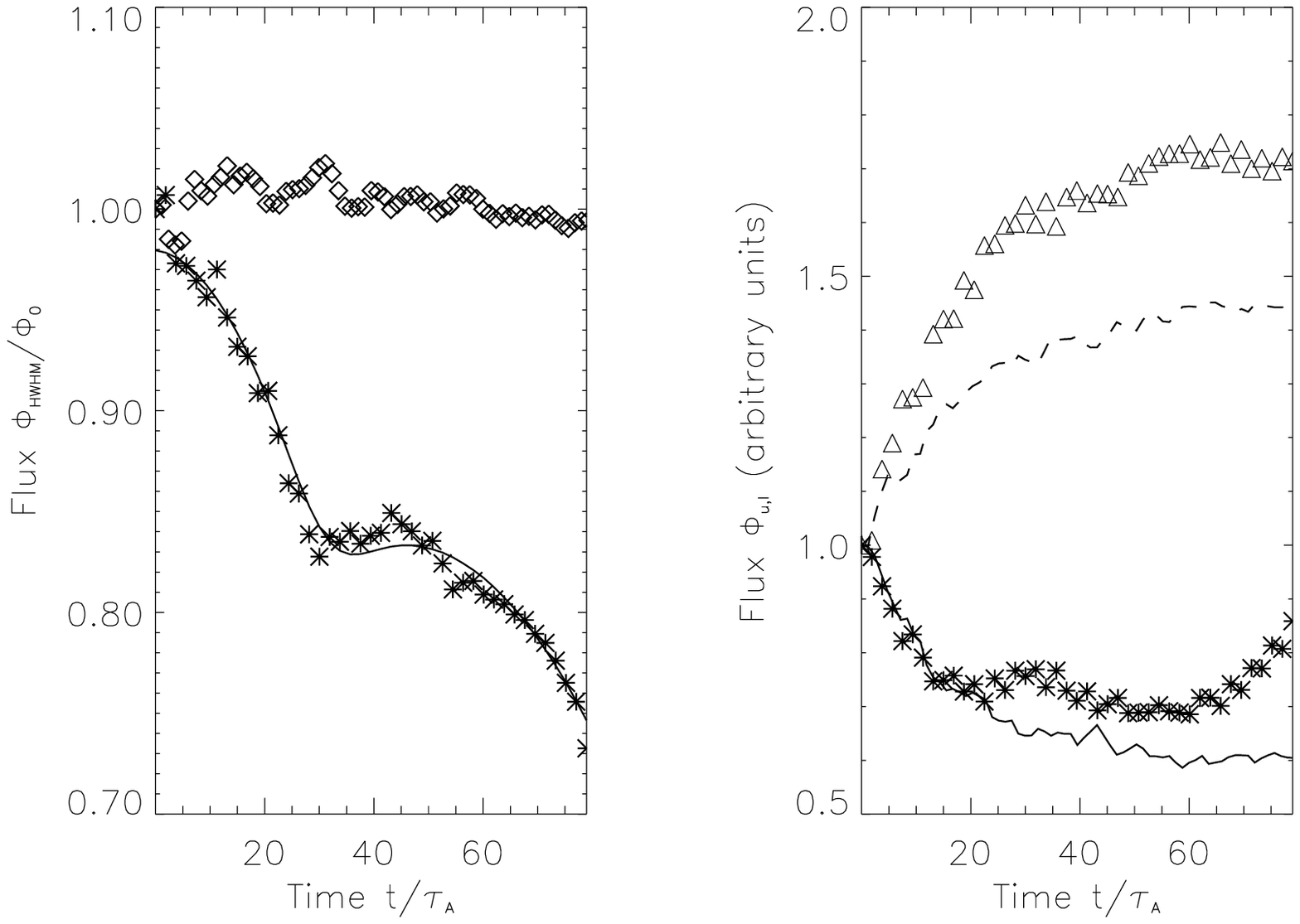} }
\caption{Left: magnetic flux within the
 rope $\Phi_i$ (3D45, stars and 2D45, diamonds)
 and an analytic fit (solid curve),
 Right: the normalized flux outside and above the center of the 3D45 rope 
 $\Phi_u$ (stars), and below, $\Phi_l$ (triangles). The same quantities are 
 shown for 2D45 
 simulation (solid and dashed curves respectively). }
\label{fig7}
\end{figure}
This is illustrated in Fig \ref{fig7}.\ (left), which shows the total 
normalized magnetic flux within the rope's HWHM-core 
$\Phi_i$ as a function of time for both the 2-d
and 3-d ropes (2D45 and 3D45). 
As the 3-d simulation progresses, the total flux-loss from 
the computational domain is only 0.3$\%$ ---
the flux content of the rope, however, decreases more rapidly. 

Also shown in Fig \ref{fig7}.\ (right) is the magnetic flux external to the 
rope $\Phi_e$ both above and below its center $\Phi_u$ and $\Phi_l$ 
respectively. Since the sum $\Phi_e + \Phi_i$ is nearly conserved,
as $\Phi_i$ decreases, $\Phi_e = \Phi_u + \Phi_l$ must increase by an equal 
amount. 
However, the distribution of the flux-loss is not symmetric around the
rope: more flux is lost 
to the surroundings below the rope than above it. 

This asymmetry also exists in 2-d, even though the 
total flux-loss is much smaller in that case. 
The asymmetry is a result of two factors. As the rope rises, the total 
volume 
above it decreases, while the volume below it increases. Furthermore, 
there is an anti-symmetry of the relative rise velocity across the rope:
when the rope ascends, there is a tendency for flux to be advected 
towards it near its apex, and transported away from it in its wake.
The more pronounced asymmetry in the 3-d case
can be attributed to the pumping effect that transports the weak
field downwards \citep{Dorch+Nordlund01}. 

We have defined the flux rope as the core magnetic structure
that lies within the HWHM-boundary. This boundary is not, however, a contour 
that moves with the fluid in the classical sense: the flux within the latter
kind of contour is naturally conserved (in ideal MHD) and
is equal to the total flux $\Phi_e + \Phi_i$. 
The HWHM-boundary is a convenient way of defining the flux rope and a
characteristic size ${\rm R}_{\rm hwhm}$, that behaves as it is expected to.
The evolution of the flux within the rope's core $\Phi_i$ is determined by
the integral of $\Delta {\bf v}\times {\bf B}$ along the rope's boundary,
where $\Delta {\bf v}$ is the difference between the fluid
velocity and the motion of the HWHM-boundary: 
the average ``slip'' $\Delta v$ is only a small fraction of the rise speed.

Making the rather crude
assumptions that the boundary only moves radially relative to the fluid
and that the circumference of the boundary is circular (which it is 
certainly not, but see Fig.\ \ref{fig3}),
the flux-loss becomes
\begin{equation} 
 \dot{\Phi}_i = - \pi {\rm R}_{\rm hwhm}~ \Delta v~ {\rm B}_c, \label{dotflux2}
\end{equation} 
where ${\rm B}_c$ is the field strength at the center of the flux rope.
Integrating the equation numerically with the quantities
determined from the simulation 3D45 the result is a
good fit to the actual flux-loss, see Fig \ref{fig7}.\ (left)---if 
$\Delta v$ is set to 3~10$^{-4}~ v_{\rm A0}$ throughout the time span
of the simulation
except for a short interval of $\sim 3~ \tau_A$ around 
$t = 30~ \tau_{\rm A}$, where $\Delta v$ changes sign,
as the rope passes from one updraft to another \citep{Dorch+ea01}.

Finally, I do not 
find flux-loss via the type of enhanced diffusion in the reference
simulations as proposed
by \cite{Petrovay+Moreno97}, and
rather, the flux-loss found here
is entirely due to the advection of flux away from the core of the flux rope
by convective motions. The turbulent erosion mechanism requires the 
turbulence to be resolved down to scales much below what I have implemented
here, i.e.\ my results does not contradict the existence of the former
effect. Most of the lost flux ends up in
the rope's wake and some is mixed back into the upper layers.
I speculate that both types of flux-loss may take place 
in the Sun, and as a result, the amount of toroidal 
flux stored near the bottom of the solar CZ may currently be underestimated.

\section{Discussion \& Summary}

In this review I have tried to shed light on some of the ``big'' questions
in the theory of buoyant magnetic flux tubes. 
Below I summarize some of the answers (A's) to the questions posed in the 
Introduction. 

~\\
{{\bf A1:} The critical pitch angle is about $30^{\rm o}$ (for thick ropes), 
 in the general
 3-d case including (large-scale) convective flows, but without rotational
 effects. The result of \cite{Abbett+ea00} that the inclusion of rotation
 lowers the critical pitch angle relative to the 2-d result, comes from 
 non-convective simulations; i.e.\ to finally answer this Q we need
 a simulation including both rotation and realistic convection.}\\
{{\bf A2:} I found no evidence that the ropes kink as they rise, at least if 
 their initial twist is low enough: one may speculate that an initially
 R-T unstable rope may stabilize due to the increase of the twist (which 
 increases only slightly in the less symmetric 3-d models).}\\
{{\bf A3:} There may be other instabilities associated with the motion through
 the uppermost strongly stratified super-adiabatic layers, but it is not know 
 from the models presented here.}\\ 
{{\bf A4:} Besides the shuffling of the flux ropes by the convective flows,
 the primary interaction is the cause of a flux-loss due to advective
 erosion.}\\
{{\bf A5:} The weak magnetic field and the flux lost from the rising flux
 tubes are transported downwards by the flux pumping-effect.}

A note on A4:
The numerical simulations show that the interaction of a buoyant twisted
flux rope with stratified convection leads to a magnetic flux-loss
from the core of the rope. 
During the simulation, the flux rope rises 96 Mm, and 
loses about 25\% of its original flux content. 
This, ceteris paribus, leads to a small increase in the 
amount of toroidal flux that must be stored at the bottom 
of the CZ during the course of the solar cycle:
Solar toroidal flux ropes rise about 200 Mm before
emerging as bipolar active regions. One may thus expect them to lose even
more of their initial flux, which would then be pumped back towards the
bottom of the CZ. Moreover,
the relative slip does not remain constant throughout the rope's rise
\citep{Dorch+ea01}. 

Of course a lot of Q's still remain to be answered (three from my list):
the most fundamental problems remaining are those of the origin of the twist, 
and the question of how it arises, Q8. This is not addressed by any of 
the models discussed here, but in my view one likely process is the 
generation of twisted field lines in large-scale flux bundles located near
the bottom of the convection zone, connecting across the solar equator: such
flux bundles would experience a rotating motion since their lower parts are
located in a region rotating slower than their uppermost parts. This rotation
would transmit a twist to the parts of the flux bundle a slightly higher
latitudes, thereby possibly giving rise to a twisted toroidal flux system.

\section*{Acknowledgments}
The author thanks the EC for support through a TMR grant to the European 
Solar Magnetometry Network. Computing time was provided by the Swedish 
National Allocations Committee. The author would also like to thank people
who helped prepare this manuscript (by lending me their figures etc.): 
Thierry Emonet, Fernando Moreno-Insertis, Tetsuya Magara, and Bill Abbett.

\section*{Appendix: Set-up of reference models}

The 2-d reference simulations lack convection and the flux tubes move in 
a polytropic atmosphere calculated by using the mass density $\rho(x)$ and 
the average gas pressure $P(x)$ in a 3-d hydrodynamic convection model via
$P=P_0 (\rho/\rho_0)^{\gamma}$, where $P_0$ and $\rho_0$ are the 
average quantities at the initial position of the flux tube. In the 3-d
reference models the full 3-d convection model is employed.

The initial set-up of the 3-d models are twofold, consisting of a snapshot
of a solar-like CZ, and of an idealized twisted magnetic flux rope. Note
that in the 2-d reference models, the CZ is polytropic through-out.

The full MHD-equations are solved
on a grid of 150 vertical times $105^2$ horizontal grid points, 
using the computational method by Galsgaard and others 
(Galsgaard \& Nordlund 1997, and Nordlund, Galsgaard \& Stein 1994):
Typical magnetic Reynolds numbers ${\rm Re}_{\rm m}$ in non-smooth regions
of the domain
are of the order of a few hundred, while
several orders of magnitude higher in smooth large-scale regions.

To model solar-like convection without including all the layers up 
to the actual surface, I use  
a simple expression for an isothermal cooling layer
at the upper boundary of the model,
restricting the effect to a thin layer. This layer is, however, 
far below the real boundary of the solar CZ. 
Horizontally the boundaries are periodic.

The hydrodynamic part of the initial condition
is a snapshot from a well developed stage of a numerical ``toy model'' of 
deep solar-like convection with a gas pressure contrast of 
roughly 2.5 orders of magnitude in the CZ alone.
The physical size of the computational box
is 250 Mm in the horizontal direction and 313 Mm in the vertical.
A cellular granulation pattern is generated on the CZ's surface
with a typical length scale of about 50 Mm,
about twice the canonical size of solar supergranules. 
The typical velocity
is about 200 m/s in the narrow downdrafts at the surface and slightly less 
in the larger upwelling regions, close to what is found for solar 
supergranulation.

Initially the entropy in the interior of the tube is set equal to that in 
the external medium.  This corresponds to buoyancy a factor of $1/\gamma$ (with 
$\gamma = 5/3$) lower than
in the case of temperature balance where the buoyancy has the
classical value of $1/\beta$. The initial twist of the flux tubes is given by
Eq.\ (\ref{initial.eq}).
The coordinate system is chosen so that $x$ is the vertical
coordinate and $z$ the coordinate along the axis which initially
is parallel to the rope.
The position of the rope is described by the set $(x_c,y_c)$, initially
equal to $(x_0,y_0)$, 
where the points $(x_c(z),y_c(z))$ along the $z$-axis are
the positions in the $(x,y)$ plane, where ${\rm B}_z$ is maximum for 
a given $z$-value.

From a 
computational point of view one  have to set $\beta_0$ lower
than the solar values ($10^4$ -- $10^7$) to reduce the computational 
time-scale to a reasonable value. 
As a compromise I set the initial field strength
so that $\beta_0 = 100$ yielding $e_{\rm M}/e_{\rm K}=100$
with the present convection model.


\begin{thebibliography}{}

\bibitem[Abbett et~al.(2000)]{Abbett+ea00}
Abbett, W.P., Fisher, G.H., Fan, Y., 2000, ApJ 540, 548

\bibitem[Abbett et~al.(2001)]{Abbett+ea01}
Abbett, W.P., Fisher, G.H., Fan, Y., 2001, ApJ 546, 1194 

\bibitem[Caligari et~al.(1995)]{Caligari+ea95} 
Caligari, P., Moreno-Insertis, F., \& Sch\"{u}ssler, M., 1995, 
ApJ 441, 886

\bibitem[Canfield et~al.(1999)]{Canfield+ea99} 
Canfield, R.C., Hudson, H.S., McKenzie, D.E., 1999, 
Geophys.\ Rev.\ Lett.\ 26(6), 627

\bibitem[D'Silva \& Choudhuri(1993)]{DSilva+Choudhuri93}
D'Silva, S., Choudhuri, A.R., 1993, A\&A 272, 621 

\bibitem[Dorch et~al.(1999)]{Dorch+ea99}
Dorch, S.B.F., Archontis, V., Nordlund, {\AA}., 1999, A\&A 352, L79 

\bibitem[Dorch et~al.(2001)]{Dorch+ea01}
Dorch, S.B.F., Gudiksen, B.V., Abbett, W.P., Nordlund, {\AA}., 2001, 
A\&A 380, 734 

\bibitem[Dorch \& Nordlund(1998)]{Dorch+Nordlund98}
Dorch, S.B.F., Nordlund, {\AA}., 1998, A\&A 338, 329

\bibitem[Dorch \& Nordlund(2001)]{Dorch+Nordlund01}
Dorch, S.B.F.,  Nordlund, {\AA}., 2001, A\&A 365, 562 

\bibitem[Emonet \& Moreno-Insertis(1998)]{Emonet+Moreno98}
Emonet, T.,  Moreno-Insertis, F., 1998, ApJ 492, 804

\bibitem[Emonet et~al.(2001)]{Emonet+ea01}
Emonet, T.,  Moreno-Insertis, F., Rast, M.P., 2001, ApJ 549, 1212 

\bibitem[Fan(2001)]{Fan2001}
Fan, Y., 2001, ApJ 554, L111 

\bibitem[Fan et~al.(1994)]{Fan+ea94}
Fan, Y., Fisher, G.H., McClymont, A.N., 1994, ApJ 436, 907

\bibitem[Fan et~al.(1999)]{Fan+ea99}
Fan, Y., Zweibel, E.G., Linton, M.G., Fisher, G.H., 1999, ApJ 521, 460 

\bibitem[Galsgaard \&  Nordlund(1997)]{Galsgaard+Nordlund97}
Galsgaard, K., Nordlund, {\AA}., 1997, Journ.\ Geoph.\ Res. 102, 219

\bibitem[Krall et~al.(1998)]{Krall+ea98}
Krall, J., Chen, J., Santoro, R., Spicer, D.S., Zalesak, S.T., Cargill, P.J., 
1998, ApJ 500, 992

\bibitem[Linton et~al.(1998)]{Linton+ea98}
Linton M.G., Dahlburg, R.B., Fisher, G.H., Longcope, D.W., 1998,
ApJ 507, 404

\bibitem[Magara \& Longcope (2001)]{Magara+Longcope01}
Magara, T., Longcope, D.W., 2001, ApJ 559, L55 

\bibitem[Matsumoto et~al.(1998)]{Matsumoto+ea98}
Matsumoto, R., Tajima, T., Chou, W., Okubo, A., Shibata, K., 1998, 
ApJ 493, L43 

\bibitem[Nordlund et~al.(1994)]{Nordlund+ea94}
Nordlund, {\AA}., Galsgaard, K., Stein, R.F., 1994,
In R.J. Rutten, C.J. Schrijver (eds.), Solar Surface Magnetic Fields
NATO ASI Series 433

\bibitem[Petrovay \&  Moreno-Insertis(1997)]{Petrovay+Moreno97}
Petrovay, K., Moreno-Insertis, F., 1997, ApJ 485, 398


\bibitem[Sch\"{u}ssler (1979)]{Schussler1979}
Sch\"{u}ssler, M., 1979, A\&A 71, 79 

\bibitem[Spruit(1981)]{Spruit1981}
Spruit, H.C., 1981, A\&A 98, 155

\bibitem[Spruit \& van Ballegooijen(1982)]{Spruit+Ballegooijen82}
Spruit, H.C., van Ballegooijen, A.A., 1982, A\&A 106, 58 

\bibitem[Sterling et~al.(2000)]{Sterling+ea00}
Sterling, A.C., Hudson, H.S., Thompson, B.J., Zarro, D.M., 2000, ApJ 532, 628 

\bibitem[Tsinganos (1980)]{Tsinganos1980}
Tsinganos, K.,  1980, ApJ 239, 746 

\bibitem[Wissink et~al.(2000)]{Wissink+ea00}
Wissink, J.G., Matthews, P.C., Hughes, D.W., Proctor, M.R.E., 2000,
ApJ 536, 982

\end{thebibliography}
\end{document}